\documentclass[conference]{IEEEtran}
\usepackage{cite}
\usepackage{amsmath,amssymb,amsfonts}
\usepackage{algorithmic}
\usepackage{graphicx}
\usepackage{textcomp}
\usepackage{xcolor}
\usepackage{stfloats}
\usepackage{amsthm}
\usepackage{subfigure}
\usepackage{enumitem}
\usepackage{arydshln}
\usepackage{cite}
\newtheorem{theorem}{Theorem}
\newtheorem{lemma}{Lemma}
\def\BibTeX{{\rm B\kern-.05em{\sc i\kern-.025em b}\kern-.08em
    T\kern-.1667em\lower.7ex\hbox{E}\kern-.125emX}}

\IEEEoverridecommandlockouts
\begin{document}
\bibliographystyle{IEEEtran}

\title{The Capacity of Multi-round Private Information Retrieval from Byzantine Databases}

\author{%
   \IEEEauthorblockN{Xinyu Yao\IEEEauthorrefmark{1},
                     Nan Liu\IEEEauthorrefmark{2}, Wei Kang\IEEEauthorrefmark{1}}
   \IEEEauthorblockA{\IEEEauthorrefmark{1}%
                     School of Information Science and Engineering, Southeast University, Nanjing, China, 
                      \{xyyao,wkang\}@seu.edu.cn}
   \IEEEauthorblockA{\IEEEauthorrefmark{2}%
                     National Mobile Communications Research Laboratory,
Southeast University, China, 
                     nanliu@seu.edu.cn}
}
\maketitle
\begin{abstract}
In this work, we investigate the capacity of private information retrieval (PIR) from $N$ replicated databases, where a subset of the databases are untrustworthy (byzantine) in their answers to the query of the user. We allow for multi-round queries and demonstrate that the identities of the byzantine databases can be determined with a small additional download cost.  As a result, the capacity of the multi-round PIR with byzantine databases (BPIR) reaches that of the robust PIR problem when the number of byzantine databases is less than the number of trustworthy databases.
\end{abstract}


\section{Introduction}
The problem of preserving the identity of the data the user retrieved from the databases is called the private information retrieval (PIR) problem, the concept of which was introduced by Chor et. al. \cite{chor1995private}. 
In the PIR problem in \cite{chor1995private}, the user wants to retrieve \emph{a certain bit} from $N$ replicated databases without revealing which bit is of interest to any single database. The main objective is to optimize the communication efficiency, which includes minimizing the upload cost and the download cost. The problem was reformulated in \cite{sun2017capacity} from an information-theoretic perspective, where  the user wants to retrieve a \emph{sufficiently large} message from the database so that the download cost is minimized. This problem was fully solved by Sun and Jafar in \cite{sun2017capacity},  where the capacity of the PIR problem was shown to be
\begin{align}
C_{\text{PIR}}= \frac{1-\frac{1}{N}}{1-(\frac{1}{N})^{K}},   \label{NanIn03}
\end{align}
which is defined as the ratio of the desired message size and the total number of downloaded symbols from the databases.  
The capacity increases with the number of databases $N$, since with the help of more databases, we can hide the privacy of the user better from any single database. 

Many interesting extensions and variations for the PIR problem have since then been studied \cite{zhang2017general, banawan2018capacity, kumar2017private, sun2018private, tajeddine2018private, lin2018mds, sun2018capacity2, tandon2017capacity, wei2018fundamental, wei2018cache, chen2017capacity, heidarzadeh2018capacity, li2018single, banawan2018multi, shariatpanahi2018multi, banawan2018asymmetry, lin2018asymmetry, banawan2018noisy, wang2017linear, banawan2018private, tian2018capacity, tajeddine2018private2, d2018lifting, tajeddine2017private, sun2017optimal, kim2017cache, abdul2017private, wang2018secure, yang2018private, jia2018cross, kumar2018private, raviv2018private}.
While most works are based on the assumption that the databases provide the user with the \emph{correct} answers  to the user's query,  
some consider the case where not all of the databases are compliant, and furthermore, the set of databases that are not compliant is not known to the user when sending its query. The first set of problems in this category is the robust PIR (RPIR) problem \cite{sun2018capacity}, where a subset of databases with size $B$ are silent and do not respond to the queries. This may be due to the fact that these databases intentionally do not respond or are slow in responding\cite{tajeddine2017robust,bitar2018staircase}. 
The goal of the user is to design its query such that the download cost is minimized, the privacy of the user is preserved, and most importantly, the desired message can be correctly decoded based only on the answers of those databases that do respond  
in a timely fashion.
The capacity of the RPIR problem is \cite{sun2018capacity}
\begin{align}
C_{\text{RPIR}}=\left\{
\begin{array}{ll}
 \frac{1-\frac{1}{N-B}}{1-(\frac{1}{N-B})^{K}} &  \text { if } B<N-1 \\
K^{-1} & \text { if } B=N-1 \\
 0 & \text{ if }B=N
 \end{array}
 \right. .
  \label{NanIn01}
\end{align} 
Comparing (\ref{NanIn01}) with (\ref{NanIn03}), we see that the capacity of the RPIR is as if the number of databases is reduced from $N$ to $N-B$, i.e., the \emph{effective number} of databases is $N-B$. Variants of the RPIR problem has been studied for coded databases \cite{tajeddine2017robust, zhang2017private}, and universal scenarios \cite{bitar2018staircase}.

The second set of problems in this category is the PIR problem from byzantine databases (BPIR). The information-theoretic formulation of the BPIR problem  was proposed in \cite{banawan2018capacity2}, where a subset of databases with size $B$ may introduce arbitrary errors to the answers of the user's query. 
This may be done unintentionally, for example, when some databases' contents are not up to date\cite{fanti2015efficient}, or intentionally, when some databases introduce errors in their answers to prevent the user from correctly decoding the desired message. While the user knows the number $B$, it does not know which set of $B$ databases are byzantine. 
The goal of the user is to design queries such that the download cost is minimized, the privacy of the user is preserved, and most importantly, the user can decode the desired message correctly despite the arbitrary wrong answers from the byzantine databases. The capacity of the BPIR problem was found in \cite{banawan2018capacity2} and is given by
\begin{align}
C_{\text{BPIR}}=\left\{
\begin{array}{ll}
\frac{N-2B}{N}\frac{1-\frac{1}{N-2B}}{1-(\frac{1}{N-2B})^{K}}  & \text { if } 2B+1<N \\
\left(NK\right)^{-1} & \text{ if }2B+1=N\\
 0 & \text{ if }2B+1>N
 \end{array}
 \right. .
\label{NanIn02}
\end{align}
Comparing (\ref{NanIn02}) with (\ref{NanIn03}) and (\ref{NanIn01}), we see that, 
the number of effective databases is $N-2B$, i.e., even though there are only $B$ byzantine databases, it is as if $2B$ databases are offering no information to the user. Note that the penalty term $\frac{N-2B}{N}$ comes from the fact that all $N$ databases send answers, though only the answers from $N-2B$ databases are useful. We do not have the penalty term of $\frac{N-B}{N}$ in the RPIR problem because the silent databases do not send any data. 
Variantions of the BPIR problem, which include collusion and coded storage have been studied in \cite{zhang2017private,tajeddine2018robust}.

There is a connection between the RPIR problem and the BPIR problem, in that if the user knows the identity of the $B$ databases who do not respond truthfully, the user can simply ignore the answers from these databases and the problem becomes the RPIR problem. 
%
Thus, if the user can be given ``some help'' in identifying the set of byzantine databases, the number of effective databases would increase from $N-2B$ to $N-B$. This help does not need to be much, compared to the download cost, as there are only ${N \choose B}$ possibilities for the set of byzantine databases, while the download cost of a sufficiently large message scales linearly with the message length. 

Based on this idea, \cite{wang2018epsilon} formulated a variant of the BPIR problem where the databases offer this ``little help'' to the user. More specifically, in their problem formulation, the databases are all trustworthy,  and there is a byzantine third party, who can listen in on the communication between $E$ databases and the user, and has the ability to arbitrarily change the answers of $B$ databases. In this case, 
as long as the databases can hide some transmitted information to the user from the adversary  and $\epsilon$-error is allowed \cite{wang2018epsilon}, the byzantine databases may be identified and the capacity of the BPIR problem reaches that of the RPIR problem, i.e., the number of effective databases is $N-B$, rather than $N-2B$. 

In this paper, we follow the problem formulation of \cite{banawan2018capacity2}, i.e., a subset of the databases are byzantine and not a third party, and explore how the user may identify the byzantine databases on its own. We
remove the assumption of single-round communication between the user and the databases in \cite{banawan2018capacity2}, and allow for multi-rounds of queries in the sense that the current round of queries can depend on the answers from the databases in the previous rounds. In the proposed achievability scheme, while the first round deals with the basic file transmission, further rounds are aimed at finding the identities of the byzantine databases. At least one byzantine database will be caught in each round, so at most $B+1$ rounds are needed. We find the capacity of the multi-round BPIR problem and show that by allowing multi-rounds, we can indeed increase the effective number of databases from $N-2B$ to $N-B$, and thus, decrease the download cost significantly. This is in contrast with the classic PIR problem, where  multi-rounds does not increase the capacity \cite{sun2018multiround}. 

\section{System Model} \label{SecSys}
Consider the problem where $K$ messages are stored on $N$ replicated databases. The $K$ messages, denoted as $W_1, W_2, \cdots, W_K$, are independent and each message consists of $L$ symbols, which are independently and uniformly distributed over a finite field $\mathbb{F}_q$, where $q$ is the size of the field, i.e.,
\begin{align}
H(W_k)&=L, \qquad k=1,...,K, \nonumber\\
H(W_1,...,W_K)&=H(W_1)+H(W_2)+\cdots +H(W_K). \nonumber
\end{align}

A user wants to retrieve the desired message $W_\theta$,  $\theta \in [K]$, by sending designed queries to the databases. Unlike in most of the previous PIR literature, where the queries are designed and fixed prior to receiving any answers from the databases, here we consider the scenario where the queries are allowed to be \emph{multi-round}, which means that the user can design the queries based on the databases' responses in the previous rounds.

More specifically, in the case where $W_\theta$ is the interested message, the query sent to the $n$-th databese in Round $m$ is denoted as $Q_{n,m}^{[\theta]}$, $n \in [N], \theta \in [K]$ and $m \in [M]$, where $M$ is the total number of rounds. Since the queries may only depend on the answers from the databases in the previous rounds, we have  
\begin{align}
I(W_{1:K}; Q_{1:N,m}^{[\theta]}|A_{1:N,1:m-1}^{[\theta]})=0, \quad \forall \theta \in [K], m \in [M]. \nonumber
\end{align}
Database $n$ in Round $m$, upon receiving the query $Q_{n,m}^{[\theta]}$, calculates the \emph{correct} answer, denoted as $\bar{A}_{n,m}^{[\theta]}$, based on the queries received in this round $Q_{n,m}^{[\theta]}$ and the messages $W_{1:K}$. Thus, we have 
\begin{align}
H(\bar{A}_{n,m}^{[\theta]}|Q_{n, m}^{[\theta]},W_{1:K})=0, \quad \forall n \in [N], m \in [M], \theta \in [K]. \nonumber
\end{align}
In the BPIR setting considered in this paper,  there exists a set of byzantine databases $\mathcal{B}$, where $|\mathcal{B}|=B$, who are untrustworthy. The remaining databases in $[N] \setminus \mathcal{B}$
are trustworthy. Hence, the trustworthy databases will transmit to the user the correct answer $A_{n,m}^{[\theta]}=\bar{A}_{n,m}^{[\theta]}$, for $n \in [N] \setminus \mathcal{B}$, while the byzantine databases will replace the correct answer $\bar{A}_{n,m}^{[\theta]}$ with an arbitrary deterministic sequence $\tilde{a}_{n,m}^{[\theta]}$ of the same size, and send it back to the user, i.e., $A_{n,m}^{[\theta]}=\tilde{a}_{n,m}^{[\theta]}$, $n \in \mathcal{B}$. 

The queries need to be designed such that the user is able to reconstruct the desired message $W_\theta$ 
after $M$ rounds no matter what arbitrary answers the byzantine databases provide, i.e., for any $\tilde{a}_{\mathcal{B},1:M}^{[\theta]}$, we have
\begin{align}
H(W_\theta|A_{1:N ,1:M}^{[\theta]},Q_{1:N, 1:M}^{[\theta]})=0,  \quad \forall \theta \in [K]\nonumber.
\end{align}

To protect the privacy of the user, we require that $\forall n \in [N]$, we have
\begin{align}
(Q_{n,1:M}^{[1]}, W_{1:K}) \sim (Q_{n,1:M}^{[\theta]}, W_{1:K}), \quad  \forall \theta \in [K]. \nonumber
\end{align}

The rate of the BPIR problem, denoted as $R$, is defined as the ratio between the message size $L$ and the total downloaded information from the databases in the worst case,
i.e.,
\begin{align}
R=\lim_{L \rightarrow \infty} \max_{\tilde{a}_{\mathcal{B}, 1:M}^{[\theta]}} \frac{L}{\sum_{n=1}^{N} \sum_{m=1}^{M} H(\bar{A}_{n,m}^{[\theta]})}. \nonumber
\end{align}

The capacity of the BPIR problem is $C_{\text{BPIR}}^{\text{multi}}=\sup R$ over all possible retrieval schemes.

\emph{Remark:} Since the databases can respond arbitrarily, it does not matter whether the databases in $\mathcal{B}$ coordinate or not in responding to the user. 

\section{Main Result}
\begin{figure}
\begin{minipage}{0.48\linewidth}
  \centerline{\includegraphics[width=4.5cm]{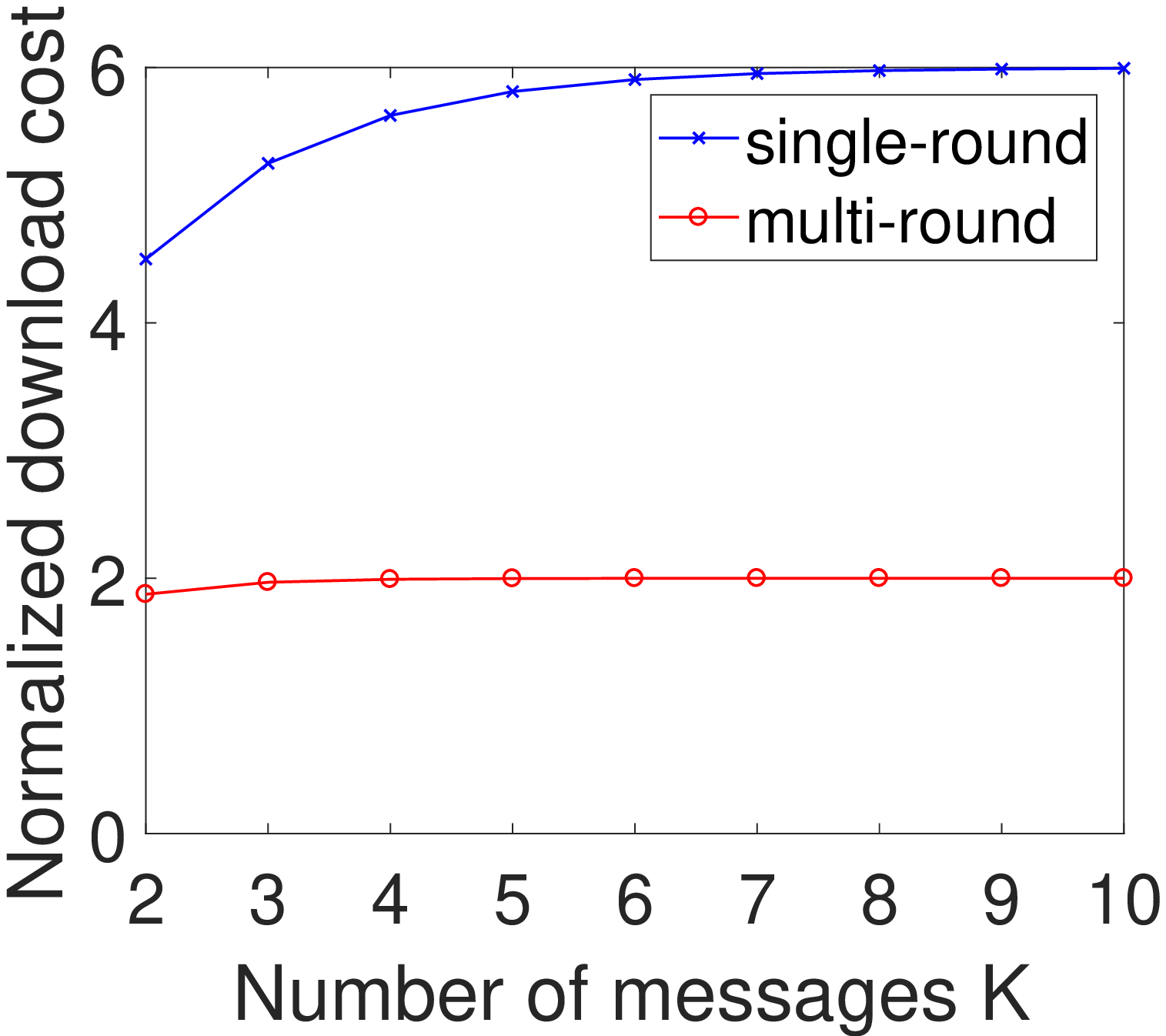}}
  \centerline{\footnotesize{(a) $N=6$, $B=2$.}}
\end{minipage}
\hfill
\begin{minipage}{.48\linewidth}
  \centerline{\includegraphics[width=4.5cm]{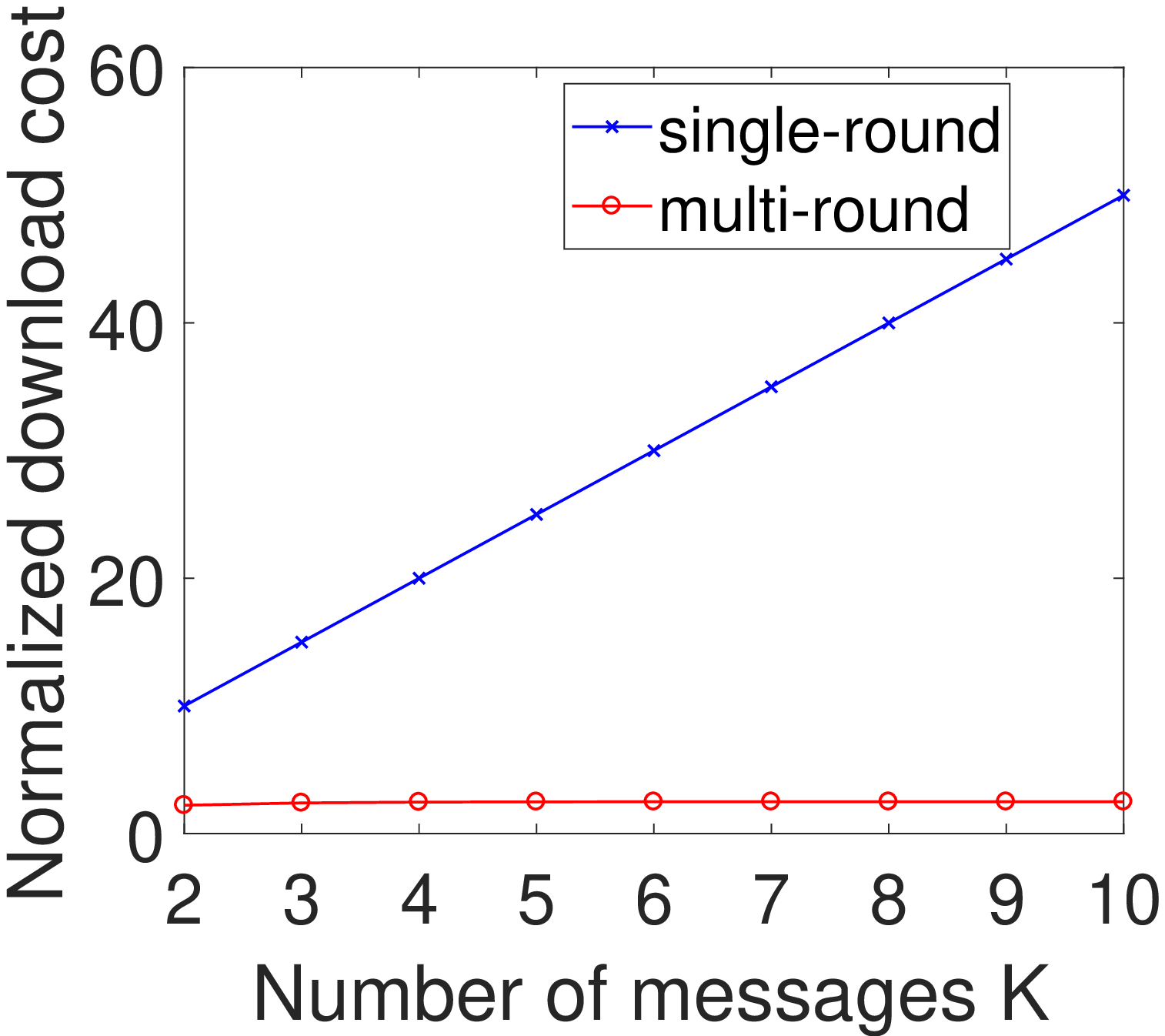}}
  \centerline{\footnotesize{(b) $N=5$, $B=2$.}}
\end{minipage}
\vfill
\caption{Download cost comparison of single-round vs. multi-round BPIR. }
\label{cmp}
\end{figure}

The main result of the paper is establishing the capacity of the multi-round BPIR problem. This is given in the following theorem. 
\begin{theorem} \label{MainResult}
The capacity of the multi-round BPIR problem is
\begin{align}
C_{\text{BPIR}}^{\text{multi}}= \left\{
\begin{array}{ll}
\frac{N-B}{N} \frac{1-\frac{1}{N-B}}{1-(\frac{1}{N-B})^{K}}, & \text{ if } 2B +1 \leq N \\
0, & \text{ if } 2B+1 > N 
\end{array} \right. .\label{NanMain01}
\end{align} 
\end{theorem}
Comparing (\ref{NanMain01}) with the capacity result for the RPIR problem, i.e., (\ref{NanIn01}), and that of the single-round BPIR problem, i.e., (\ref{NanIn02}), we see that  
\begin{enumerate}
\item When $2B+1 < N$, by allowing multi-round, the number of effective databases has increased from $N-2B$, which is the case for single-round BPIR, to $N-B$, which is the same as the RPIR problem. An example of the normalized download cost reduction is shown in Fig. \ref{cmp} (a), where the normalized download cost is defined as the inverse of the capacity of the PIR problem. 
\item When $2B+1 = N$, by allowing multi-round, the download cost is significantly reduced since in a single-round BPIR, the only possible scheme was to download all the messages from all the databases. An example of the normalized download cost reduction is shown in Fig. \ref{cmp} (b). 
\item When $2B+1 > N$, allowing multi-round does not help, as the majority of the databases are untrustworthy and there are error instances introduced by the byzantine databases where correct decoding at the user is impossible. Note that under the problem formulation of \cite{wang2018epsilon}, where the databases are all trustworthy and a third-party is performing the byzantine attack, even when $2B+1>N$, the user can still correctly decode the desired message with the databases' help, which is hidden from the byzantine third-party.
\end{enumerate}

The achievability proof of Theorem \ref{MainResult} is given in the next section, where the $B$ byzantine databases are identified by the multi-round queries. The converse proof, which is more trivial, is presented in Section V.

\section{Achievability}
We will first provide the main idea of the achievability scheme. The proposed scheme is performed in several rounds. 
In the first round, the message is cut into blocks, and for each block we use the query structure of the RPIR scheme\cite{sun2018capacity}. While the answers in this round contain arbitrary errors introduced by the byzantine databases, due to the error detecting capability of linear block codes, the user is able to detect the set of blocks that contain errors, which we call the error blocks. In the following rounds, say Round $m$, $m \geq 2$, the user re-request \emph{one} error block using an MDS code whose rate is small enough for the errors in the answers of Round $m$ to be corrected. By comparing the corrected data received in Round $m$ to the answers of this error block in Round 1, at least one of the byzantine databases will be identified. After at most $B+1$ rounds, the user can identify all  of the byzantine databases that introduced errors in Round 1. Ignoring the answers from these byzantine databases, the user can decode the desired message correctly. Since the message length is sufficiently long, which means that the number of blocks is sufficiently large, the extra download cost in the rounds after Round 1 is negligible.

The details of the proposed achievability scheme is below.
We first start with some preliminaries. 

\subsection{Some Preliminaries}
We first recall the error-detection and error-correction capabilities of linear block codes. 
\begin{lemma}(Code Capability\cite{van2012introduction}) \label{Nan01}
Let C be an $[n, k, d]$ linear block code over $\mathbb{F}_{q}$. Then the code  is able to detect up to $(d-1)$ errors and correct up to $\lfloor \frac{d-1}{2} \rfloor$ errors.
\end{lemma}

Next, recall the definition of the MDS (maximum distance separable) code \cite{ling2004coding}. A linear code with parameters $[n, k, d]$ such that $k+d = n+1$ is called an MDS code. The existence of the MDS code is given by \cite{macwilliams1977theory}, which states that if the field size $q$ is large enough, then there exists an ${[n,k]}_{q}$ MDS code for any $k<n$.

\begin{lemma}(Punctrued MDS Code\cite{feyling1993punctured}) \label{puncture}
Let C be an (n,k) MDS code, then deleting any $p<n-k$ symbols from the codeword yields a punctured code, which is also an MDS code.
\end{lemma}

Before presenting the detailed achievability scheme, 
We first provide a motivating example.

\subsection{Example: $N=6, K=2, B=2$.}
Consider the case of retrieving with privacy one of two messages from six databases. Two of the six databases are byzantine. The message length is $L = (N-B)^{K} l = 16 l$.  Each message is cut into $l$ blocks. Let $x_{1,i}^{j}$ and $x_{2,i}^{j}$ denote the $i$-th symbol in block $j$ of $W_1$ and $W_2$, respectively. Assume without loss of generality that $W_1$ is the desired message.

\textbf{Transmission in Round 1:} in Round 1, 
the user constructs the queries in the same way for each block. For block $j$, the query to retrieve the 16 symbols $x^j_{1,[1:16]}$ is constructed in the following way. Collect the 16 symbols into a column vector $\mathbf{x}_{1}^{j}=\begin{bmatrix}x_{1,1}^{j} &x_{1,2}^{j} &\cdots &x_{1,16}^{j}\end{bmatrix}^T$. Let $S_1, S_2 \in \mathbb{F}_{q}^{16\times 16}$ be random matrices chosen privately by the user, independently and uniformly from all $16\times 16$ full-rank matrices over $\mathbb{F}_q$. For the desired message $W_{1}$, $S_{1}$ is used to create 16 linear combinations of $x_{1,[1:16]}^{j}$. Then these 16 mixed symbols are further mapped into 24 symbols $a_{[1:24]}^{j}$ using a (24,16)-MDS code, i.e., 
\begin{align}
a_{[1:24]}^{j} \triangleq \mathbf{MDS}_{24\times 16} S_{1} \mathbf{x}_{1}^{j}. \label{Nan03}
\end{align}

For the undesired message $W_2$, consider the first 4 rows of the random matrix $S_{2}$. The first 4 rows of $S_{2}$ maps the 16 undesired symbols $x_{2,[1:16]}^{j}$ into 4 linear combinations. They are further mapped into 24 symbols, denoted as $b_{[1:24]}^{j}$, with a (24,4)-MDS code, i.e.
\begin{align}
b_{[1:24]}^{j} \triangleq \mathbf{MDS}_{24 \times 4} S_{2}([1:4],:) \mathbf{x}_{2}^{j}. \label{Nan02}
\end{align}
We note here that 
the random matrices $S_1,S_2$ and the MDS code is the same for each block. 

As shown in Table \ref{tab1}, the first stage of the query structure is to retrieve single symbols. The second stage involves using the undesired message as side information to receive 2-sum symbols. Since there are two byzantine databases, three independent undesired symbols from another three databases can be utilized in the second stage for each database.

\begin{table}[htbp]
\caption{The query table for Block $j$ in Round 1 for the case of $N=6, K=2, B=2$}
\begin{center}
\scalebox{0.9}{
\begin{tabular}{|c|c|c|c|c|c|}
 \hline
DB1 & DB2 & DB3 & DB4 & DB5 & DB6\\
 \hline
$a^j_1$ & $a^j_2$ & $a^j_3$ & $a^j_4$ & $a^j_5$ & $a^j_6$\\
$b^j_1$ & $b^j_2$ & $b^j_3$ & $b^j_4$ & $b^j_5$ & $b^j_6$\\
 \hline
$a^j_{7}+b^j_{7}$ & $a^j_{8}+b^j_{8}$ & $a^j_{9}+b^j_{9}$ & $a^j_{10}+b^j_{10}$ & $a^j_{11}+b^j_{11}$ & $a^j_{12}+b^j_{12}$\\
$a^j_{13}+b^j_{13}$ & $a^j_{14}+b^j_{14}$ & $a^j_{15}+b^j_{15}$ & $a^j_{16}+b^j_{16}$ & $a^j_{17}+b^j_{17}$ & $a^j_{18}+b^j_{18}$\\
$a^j_{19}+b^j_{19}$ & $a^j_{20}+b^j_{20}$ & $a^j_{21}+b^j_{21}$ & $a^j_{22}+b^j_{22}$ & $a^j_{23}+b^j_{23}$ & $a^j_{24}+b^j_{24}$\\
 \hline
\end{tabular}}
\label{tab1}
\end{center}
\end{table}
Round 1  ends when the answers for all $l$ blocks are received by the user. 

Note that the scheme described above for each block  is the same as the query scheme of the RPIR problem for the entire message \cite{sun2018capacity}. Similar to the case of RPIR, 
the transmission of Block $j$ is private, and so is the transmission of the entire Round 1.

\textbf{Analyzing received data after Round 1:} the user examines the received data as follows. 
%
For Block $j$, the user first looks at the retrieved single symbols of the undesired message, i.e., $b_{[1:6]}^{j}$. $b_{[1:6]}^{j}$ is a (24,4)-MDS code with a sequence of length $p=18$ removed. Since the removed length $p<24-4$, according to Lemma \ref{puncture}, $b_{[1:6]}^{j}$ is a $(6,4)$-MDS code, which can detect up to two errors, according to Lemma \ref{Nan01}. Note that the maximum number of errors the byzantine databases can introduce in $b_{[1:6]}^{j}$ is 2. Hence, if one or both of the byzantine databases introduced an error in this stage, the user would be able to detect it. If errors are detected in $b_{[1:6]}^{j}$, Block $j$ is declared an \emph{error block of Round 1}. Otherwise, $S_{2}([1:4],:) \mathbf{x}_{2}^{j}$ is correctly decoded. 

If no error was found in $b_{[1:6]}^{j}$, 
based on the four decoded symbols $S_{2}([1:4],:) \mathbf{x}_{2}^{j}$, the user calculates $b_7^{j}$ to $b_{24}^{j}$ using (\ref{Nan02}). $b_7^{j}$ to $b_{24}^{j}$ is then subtracted from the received 2-sum symbols in Stage 2 and we are left with $a_{[7:24]}^{j}$. Since $a_{[1:24]}^{j}$ is a (24,16)-MDS code, and the number of errors the two byzantine databases can introduce is less than $8$, the user can detect if any error is introduced by the byzantine databases in $a_{[1:24]}^{j}$, according to Lemma \ref{Nan01}. If errors are detected,
Block $j$ is declared an \emph{error block of Round 1}.

We repeat the above procedure for each of the $l$ blocks.  At the end of the procedure, denote the set of error blocks
as $\mathcal{E}_1$. 
We have one of the two following cases:

\noindent
1) $\mathcal{E}_1=\phi$, i.e., no error was found on any of the blocks, which means that the byzantine databases did not attack in Round 1.
In this case, no further rounds are needed. The user proceeds to the final decoding step. 

\noindent
2) $|\mathcal{E}_1| \geq 1$, which means that the byzantine databases attacked in Round 1. 
%
In this case, further rounds are needed to correctly decode the desired message. 

\textbf{Transmission in Round 2:} the user uniformly picks one block out of the $|\mathcal{E}_1|$ blocks with detected error, say Block $j_1$, and perform the following encoding:
map the 16 symbols of Block $j_1$ of each message into 48 symbols  using a (48,16)-MDS code,
\begin{align}
c_{[1:48]}^{j_1} \triangleq \mathbf{MDS}_{48\times 16} \mathbf{x}_{1}^{j_1},  \nonumber \\
d_{[1:48]}^{j_1} \triangleq \mathbf{MDS}_{48\times 16} \mathbf{x}_{2}^{j_1}.  \nonumber
\end{align}
Ask the databases to transmit the symbols as shown in Table \ref{tab2}.

\begin{table}[htbp]
\caption{The query table for Block $j$ in Round 2 for the case of $N=6, K=2, B=2$}
\begin{center}
\begin{tabular}{|c|c|c|c|c|c|}
 \hline
DB1 & DB2 & DB3 & DB4 & DB5 & DB6\\
 \hline
$c^{j_1}_1$,$d^{j_1}_1$ & $c^{j_1}_2$,$d^{j_1}_2$ & $c^{j_1}_3$,$d^{j_1}_3$ & $c^{j_1}_4$,$d^{j_1}_4$ & $c^{j_1}_5$,$d^{j_1}_5$ & $c^{j_1}_6$,$d^{j_1}_6$\\
$c^{j_1}_{7}$,$d^{j_1}_{7}$ & $c^{j_1}_{8}$,$d^{j_1}_{8}$ & $c^{j_1}_{9}$,$d^{j_1}_{9}$ & $c^{j_1}_{10}$,$d^{j_1}_{10}$ & $c^{j_1}_{11}$,$d^{j_1}_{11}$ & $c^{j_1}_{12}$,$d^{j_1}_{12}$\\
$c^{j_1}_{13},d^{j_1}_{13}$ & $c^{j_1}_{14},d^{j_1}_{14}$ & $c^{j_1}_{15},d^{j_1}_{15}$ & $c^{j_1}_{16},d^{j_1}_{16}$ & $c^{j_1}_{17},d^{j_1}_{17}$ & $c^{j_1}_{18},d^{j_1}_{18}$\\
$c^{j_1}_{19},d^{j_1}_{19}$ & $c^{j_1}_{20},d^{j_1}_{20}$ & $c^{j_1}_{21},d^{j_1}_{21}$ & $c^{j_1}_{22},d^{j_1}_{22}$ & $c^{j_1}_{23},d^{j_1}_{23}$ & $c^{j_1}_{24},d^{j_1}_{24}$\\

$c^{j_1}_{25},d^{j_1}_{25}$ & $c^{j_1}_{26},d^{j_1}_{26}$ & $c^{j_1}_{27},d^{j_1}_{27}$ & $c^{j_1}_{28},d^{j_1}_{28}$ & $c^{j_1}_{29},d^{j_1}_{29}$ & $c^{j_1}_{30},d^{j_1}_{30}$\\
$c^{j_1}_{31},d^{j_1}_{31}$ & $c^{j_1}_{32},d^{j_1}_{32}$ & $c^{j_1}_{33},d^{j_1}_{33}$ & $c^{j_1}_{34},d^{j_1}_{34}$ & $c^{j_1}_{35},d^{j_1}_{35}$ & $c^{j_1}_{36},d^{j_1}_{36}$\\
$c^{j_1}_{37},d^{j_1}_{37}$ & $c^{j_1}_{38},d^{j_1}_{38}$ & $c^{j_1}_{39},d^{j_1}_{39}$ & $c^{j_1}_{40},d^{j_1}_{40}$ & $c^{j_1}_{41},d^{j_1}_{41}$ & $c^{j_1}_{42},d^{j_1}_{42}$\\
$c^{j_1}_{43},d^{j_1}_{43}$ & $c^{j_1}_{44},d^{j_1}_{44}$ & $c^{j_1}_{45},d^{j_1}_{45}$ & $c^{j_1}_{46},d^{j_1}_{46}$ & $c^{j_1}_{47},d^{j_1}_{47}$ & $c^{j_1}_{48},d^{j_1}_{48}$\\
 \hline
\end{tabular}
\label{tab2}
\end{center}
\end{table}

\textbf{Analyzing received data after Round 2:} 
the maximum number of errors that can be introduced by the two byzantine databases is 16 for $c_{[1:48]}^{j_1}$$(d_{[1:48]}^{j_1})$. Since the $(48,16)$-MDS code can correct up to 16 errors according to Lemma \ref{Nan01}, irrespective of the errors the two byzantine databases introduce in the answers of Round 2, upon receiving the transmission in Table \ref{tab2}, $\mathbf{x}_{1}^{j_1}$ and $\mathbf{x}_{2}^{j_1}$ can both be correctly decoded.

Now, the user calculates $a_{[1:24]}^{j_1}$ and $b_{[1:24]}^{j_1}$ according to (\ref{Nan03}) and (\ref{Nan02}) using the correctly decoded $x_{1,[1:16]}^{j_1}$ and $x_{2,[1:16]}^{j_1}$. Then, the user compares the calculated correct values of $a_{[1:24]}^{j_1}$ and $b_{[1:24]}^{j_1}$ with the answers for Block $j_1$ received from the six databases in Round 1, and the errors introduced by the byzantine databases in Round 1 can be identified. We have one of the following two cases:

\noindent
1) Both byzantine databases introduced an error in Block $j_1$. In this case, the user can identifie both byzantine databases. No further rounds are needed. The user proceeds to the final decoding step. 

\noindent
2) Only one of the byzantine databases introduced an error in Block $j_1$. In this case, the user can identify one of the byzantine databases, say Database $y_1$. For all the blocks in $\mathcal{E}_1$, i.e., the blocks detected with errors in Round 1, we ignore the answer from the byzantine database identified, i.e., Database $y_1$, and detect whether there are still errors in the blocks in $\mathcal{E}_1$.  This detection procedure is the same as that after Round 1, i.e., for each $j \in \mathcal{E}_1$, we first detect to see if the $b_{[1:y_1-1] \bigcup [y_1+1:6]}^{j}$ still has errors. Since $b_{[1:y_1-1] \bigcup [y_1+1:6]}^{j}$ is a (5,4)-MDS code, and at most 1 error is introduced by the unidentified byzantine database, the error, if introduced, will be detected. If an error is detected, we pronouce this block an \emph{error block of Round 2}. If $b_{[1:y_1-1] \bigcup [y_1+1:6]}^{j}$ has no errors, the user correctly decodes $S_{2}([1:4],:) \mathbf{x}_{2}^{j}$, and proceeds to calculate $b_7^{j}$ to $b_{24}^{j}$ using (\ref{Nan02}), which is then subtracted from the received 2-sum symbols in Stage 2 of Round 1, after which we are left with $a_{[7:24]}^{j}$. Since $a_{[1:24]}^{j}$ ignoring the answer from the identified byzantine database is a (20,16)-MDS code, and at most 4 errors can be introduced by the unidentified byzantine database, the introduced errors will  be detected. If errors are detected, Block $j$ is an \emph{error block of Round 2}. Denote the set of error blocks, i.e., blocks where errors still exist after the above detection procedure, as $\mathcal{E}_2$. 
We have one of the following two cases:

\noindent
1) $\mathcal{E}_2 =\phi$, which means that the other byzantine database never introduced any error in Round 1. In this case, no further rounds are needed. The user proceeds to the final decoding step.


\noindent
2) $|\mathcal{E}_2| \geq 1$, which means that the other uncaught byzantine database introduced errors in the blocks belonging to $\mathcal{E}_2$. We need Round 3 to catch the remaining uncaught byzantine database.

\textbf{Transmission in Round 3:} uniformly and randomly pick one of the blocks in $\mathcal{E}_2$, say Block $j_2$. Repeat the query procedure described in Round 2 for Block $j_1$, replacing $j_1$ with $j_2$. Upon receiving the $96$ answered symbols from the databases in Round 3, we may again decode the corrected symbols of $\mathbf{x}_{1}^{j_2}$ and $\mathbf{x}_{2}^{j_2}$, due to the use of the (48,16)-MDS code. By calculating $a_{[1:24]}^{j_2}$ and $b_{[1:24]}^{j_2}$  according to (\ref{Nan03}) and (\ref{Nan02}) with the correctly decoded $x_{1,[1:16]}^{j_2}$ and $x_{2,[1:16]}^{j_2}$, we may compare the calculated correct values of $a_{[1:24]}^{j_2}$ and $b_{[1:24]}^{j_2}$ with the corresponding Block $j_2$ received from the six databases in Round 1. This comparison will for sure reveal the identify of the uncaught byzantine database as it had introduced at least one error in this block. 
By now, the user has caught both Byzantine databases. No further rounds are needed. The user proceeds to the final decoding step.


\textbf{Final Decoding:} the user has identified all the byzantine databases that introduced errors in Round 1, which we call dishonest databases. We call the set of databases that did not introduce any errors in Round 1 as honest databases. 
The user  examines the data received in Round 1,
ignore the answers from the dishonest databases and decode only from the answers of the honest databases. Note that the number of honest databases is no less than 4. More specifically, for block $j$, $j \in [l]$, looking at the received symbols of $b_{[1:6]}^j$ from the honest databases, the user can correctly decode $S_{2}([1:4],:) \mathbf{x}_{2}^{j}$, based on the MDS property of the $(24,4)$-MDS code. Next, the user calculates $b_7^{j}$ to $b_{24}^{j}$ using (\ref{Nan02}). Substract $b_7^{j}$ to $b_{24}^{j}$ from the received 2-sum symbols to obtain $a_{[7:24]}^{j}$. Looking at the received symbols of $a_{[1:24]}^{j}$ from the honest databases, the user can correctly decode $x_{1, [1:16]}^j$ based on the MDS property of the $(24,16)$-MDS code. This is repeated for each block $j$, $j \in [l]$ to correctly decode the entire desired message. 


Thus, for this example, we have proposed a query scheme that allows the user to decode the desired message, irrespective of the error pattern introduced by the byzantine databases.
%
%
 It is easy to see that the scheme is private, not only in Round 1, but in the entire query process, as in retransmitting, the user uniformly and randomly picked a block with errors detected and asked for the block of both messages, thus displaying no preference over Message 1 or 2. Finally, the download cost is $30l$ if only 1 round is needed, $30l+96$ if 2 rounds are needed and $30l+192$ if 3 rounds are needed, which is the worst case. Thus, the BPIR achievable rate is
\begin{align}
\lim_{l \rightarrow \infty} \frac{16l}{30l+192}=\frac{8}{15}, \nonumber
\end{align}
consistent with Theorem \ref{MainResult}.

\subsection{General Scheme}
This scheme works when we have $2B+1 \leq N$. Suppose each message consists of $L=(N-B)^{K}\cdot l$ symbols from $\mathbb{F}_{q}$, and $W_{\theta}$ is the desired message. The user cuts the message into $l$ blocks and constructs the queries in the same way for each block. Let the $j$-th block of Message $k$ be denoted as a vector $\mathbf{x}_k^j$ of length $(N-B)^k$, $j \in [l]$, $k \in [K]$.

\textbf{Transmission in Round 1:} 
We construct the queries of each block using the achievable scheme of robust PIR for the entire message \cite[Section IV]{sun2018capacity}. In \cite{sun2018capacity}, $N$ out of $M$ databases respond and any $T$ databases may collude. For our problem, we use the query scheme of \cite[Section IV]{sun2018capacity} with $M$ replaced by $N$, $N$ replaced by $N-B$ and $T=1$. 
More specifically, the queries for each block involve $(N-B)^K$ symbols from each message. 
For the desired message $W_{\theta}$, use $(N(N-B)^{K-1},(N-B)^{K})$-MDS code to encode the symbol mixtures randomized by an uniformly chosen matrix $S_{\theta}$ from $\mathbb{F}_q^{(N-B)^{K} \times (N-B)^{K}}$. Consider the $\delta=2^{K-1}$ subsets of $[K]$ that contain $\theta$, each of which denoted as $\mathcal{L}_{i},i\in[\delta]$. Define the vector of symbol mixtures of the desired message as
\begin{align}
\begin{bmatrix}
u_{\theta,\mathcal{L}_{1}}^{j} \\
u_{\theta,\mathcal{L}_{2}}^{j} \\
\vdots \\
u_{\theta,\mathcal{L}_{\delta}}^{j} \\
\end{bmatrix}
= \mathbf{MDS}_{N(N-B)^{K-1}\times(N-B)^{K}}S_{\theta}\mathbf{x}^j_{\theta}. \label{Nan06}
\end{align}
For undesired messages $\mathbf{x}^j_{k},k\in[K]\setminus\{\theta\}$, denote each of the $\Delta=2^{K-2}$ subsets of $[K]$ that contains $k$ and not $\theta$ as $\mathcal{K}^k_i,i\in[\Delta]$. Define the vector of symbols mixtures of $\mathbf{x}^j_{k},k\in[K]\setminus\{\theta\}$, as (\ref{LongEquation}) shown on the top of the this page, 
\begin{figure*}
\begin{align}
\begin{bmatrix}
\begin{matrix}
u_{k,\mathcal{K}^k_{1}}^{j} \\
u_{k,\mathcal{K}^k_{1}\cup\{\theta\}}^{j} \\
\end{matrix} \\ \hdashline[2pt/2pt]
\begin{matrix}
u_{k,\mathcal{K}^k_{2}}^{j} \\
u_{k,\mathcal{K}^k_{2}\cup\{\theta\}}^{j}
\end{matrix} \\ \hdashline[2pt/2pt]
\begin{matrix}
\vdots
\end{matrix} \\ \hdashline[2pt/2pt]
\begin{matrix}
u_{k,\mathcal{K}^k_{\delta}}^{j} \\
u_{k,\mathcal{K}^k_{\delta}\cup\{\theta\}}^{j}
\end{matrix}
\end{bmatrix}
=
\begin{bmatrix}
\begin{matrix}
\mathbf{MDS}_{N\alpha_1\times\alpha_1} & 0 & 0 & 0 \\ \hdashline[2pt/2pt]
0 & \mathbf{MDS}_{N\alpha_2\times\alpha_2} & 0 & 0 \\ \hdashline[2pt/2pt]
0 & \cdots & \ddots & 0 \\ \hdashline[2pt/2pt]
0 & 0 & 0 & \mathbf{MDS}_{N\alpha_{\Delta}\times\alpha_{\Delta}}
\end{matrix}
\end{bmatrix}
S_{k}[(1:(N-B)^{K-1}),:]\mathbf{x}_{k}^j \label{LongEquation}
\end{align}
\end{figure*}
where $\alpha_i=(N-B)(N-B-1)^{|\mathcal{K}_i|-1}$, each $u_{k,\mathcal{K}^k_i}^{j}$ is a $\frac{N}{N-B}\alpha_i\times1$ vector, and each $u_{k,\mathcal{K}^k_i\cup\{\theta\}}^{j}$ is a $\frac{N(N-B-1)}{N-B}\alpha_i\times1$ vector.

For each non-empty subset $\mathcal{K}\subseteq [K]$, generate the query vectors as
\begin{align}
\sum_{k\in\mathcal{K}}u_{k,\mathcal{K}}^{j}. \label{Nan08}
\end{align}
Distribute the elements of the query vector evenly among the $N$ databases to complete the construction of the queries.


\textbf{Analyzing received data after Round 1:} 
After all $l$ blocks are received in Round 1, the user examines the data of each block for error. 

For Block $j$, the user first looks at the received undesired symbols. For $\mathcal{K} \subseteq [K]$ that does not include $\theta$,  
$\sum_{k \in \mathcal{K}} u_{k,\mathcal{K}}^{j}$ is an
$(\frac{N}{N-B}\alpha_i,\alpha_i)$-MDS code, which can detect up to $\frac{B}{N-B}\alpha_i$ errors, according to Lemma \ref{Nan01}.  Each database transmit $\frac{1}{N}\frac{N}{N-B}\alpha_i$ symbols, and as a result, the $B$ byzantine databases can make at most $\frac{B}{N}\frac{N}{N-B}\alpha_i$ errors. Thus  if any byzantine databases introduced errors in the downloaded symbols of $\sum_{k \in\mathcal{K}} u_{k,\mathcal{K}}^{j},\mathcal{K} \subseteq [K], \theta \notin \mathcal{K}$, the user can detect them. If errors are detected in the undesired symbols of Block $j$, we say Block $j$ is an \emph{error block of Round 1}. 
Otherwise, the interference terms in $\sum_{k \in \mathcal{K}} u_{k,\mathcal{K}}^{j}$, $\theta \in \mathcal{K}$ can be correctly calculated and subtracted.

If no error was found in the undesired symbols of Block $j$,  by subtracting the undesired symbol mixtures, the user can obtain all the desired symbols mixtures, i.e., $u_{\theta, \mathcal{K}}^j$, for all $\mathcal{K}$ that contain $\theta$. Since an $(N(N-B)^{K-1},(N-B)^{K})$-MDS code is used for the desired symbols according to (\ref{Nan06}), where at most $B(N-B)^{K-1}$ errors are introduced by the byzantine databases, if any byzantine database introduced errors in the desired symbol mixtures, the user can detect them. If errors are detected in the desired symbol mixtures of Block $j$, we say Block $j$ is an \emph{error block of Round 1}. 

We repeat the above procedure for each of the $l$ blocks.  At the end of the procedure, denote the set of error blocks of Round 1 as $\mathcal{E}_1$. Depending on the actions of the byzantine databases, we have one of the following two cases: 

\noindent
1) $\mathcal{E}_1 = \phi$, which means that no error was introduced in Round 1 by the byzantine databases. In this case, no further rounds are needed. The user proceeds to the final decoding step. 

\noindent
2) $|\mathcal{E}_1| \geq 1$, which means that some byzantine databases introduced errors in Round 1. In this case, further rounds are performed according to the following iterative precedure. At the end of Round $m-1$, $m \geq 2$, the user has found the set of error blocks of Round $m-1$, denoted as $\mathcal{E}_{m-1}$. At the beginning of Round $m$, the user uniformly and randomly picks one block, say $j_{m-1}$, out of the $|\mathcal{E}_{m-1}|$ error blocks and constructs queries for Block $j_{m-1}$ in Round $m$. 

\textbf{Transmission in Round $m$ ($m\geq2$):} 
the user downloads the whole Block $j_{m-1}$ of all $K$ messages using an $(N \alpha,(N-B)^{K})$-MDS code for each message, i.e.,
\begin{align}
u_{k, [1:N \alpha]}^{j_{m-1}} \triangleq \mathbf{MDS}_{N \alpha \times (N-B)^{K}} \mathbf{x}_{k}^{j_{m-1}}, \quad \forall k \in [K] \nonumber
\end{align}
where $\alpha \triangleq \left\lceil\frac{(N-B)^{K}+1}{N-2B} \right\rceil$. 
 By retrieving the whole block of all messages, queries in Round $m$ do not conflict with the previous rounds in terms of privacy. The maximum number of errors introduced by the byzantine databases is $B \alpha$
 for each message, which is less than $\left \lfloor \frac{d-1}{2} \right\rfloor$, where the distance $d$ of the $(N \alpha,(N-B)^{K})$-MDS code is $ N \alpha-(N-B)^{K}$. Thus, according to Lemma 1, irrespctive of the errors the byzantine databases introduce in Round $m$ for Block $j_{m-1}$, the correct symbols can be decoded by the user.

\textbf{Analyzing received data after Round $m$:} 
With the correct symbols of Block $j_{m-1}$ decoded, the user can calculate the correct answers for Block $j_{m-1}$ in Round 1 using (\ref{Nan06}), (\ref{LongEquation}) and (\ref{Nan08}), and compare it with the actual received symbols of Block $j_{m-1}$ in Round 1. Since Block $j_{m-1}$ is an error block of Round $m-1$, it means that at least one of the unidentified byzantine databases has made errors in Block $j_{m-1}$ of Round 1. Hence, with the comparison, at least one of the byzantine databases not caught in the previous rounds will be caught in this round. Denote the number of byzantine databases caught in Round $m$ as $n_m$. 

The user finds the set of error blocks of Round $m$, i.e., $\mathcal{E}_m$, as follows: for each of the error blocks of $\mathcal{E}_{m-1}$, detect if there are still errors after ignoring the answer from the byzantine databases whose identity has been discoverd thus far. This error detection can be done as we are looking at the $(\frac{N-\sum_{i=2}^{m}n_i}{N-B}\alpha_i,\alpha_i)$-MDS code for undesired symbols with at most $\frac{B-\sum_{i=2}^{m}n_i}{N-B}\alpha_i$ remaining errors, and $((N-\sum_{i=2}^{m}n_i)(N-B)^{K-1},(N-B)^{K})$-MDS code for the desired symbols with at most $(B-\sum_{i=2}^{m}n_i)(N-B)^{K-1}$ remaining errors. According to Lemma 1, these amounts of errors can be detected. Denote the set of blocks where errors still exist as $\mathcal{E}_{m}$. We have one of the following two cases:

\noindent
1) $\mathcal{E}_{m}=\phi$, which means that the user has found out the identity of all byzantine databases that introduced errors in Round 1. In this case, no further rounds are needed. The user proceeds to the final decoding step. 

\noindent
2) $|\mathcal{E}_{m}|\geq1$, which means that there are still uncaught byzantine databases who introduced errors in Round 1. In this case, the user starts the query for Round $m+1$ by repeating the above procedure. 

\textbf{Final Decoding:} We have identified all the byzantine databases that introduced errors in Round 1. The user decodes by ignoring their answers. Since the query structure of Round 1 resembles that of the RPIR where $N-B$ databases respond, we can correctly decode the desired message ignoring the answers from the byzantine databases.

Now we calculate the achievable rate. 
The downloaded symbols for each block in Round 1 is $N \cdot \sum_{k=1}^{K}(N-B-1)^{k-1}{K \choose k}$, according to the RPIR problem \cite[Section IV]{sun2018capacity}. Starting from Round $m$, $m \geq 2$, $KN\alpha$ symbols are downloaded in each round.
In the worst case, the user needs $B$ rounds to catch all the $B$ byzantine databases. Hence, the achievable rate of the multi-round BPIR scheme proposed is
 \begin{align}
R \nonumber & = \lim_{l \rightarrow \infty} \frac{(N-B)^{K}l}{N \sum_{k=1}^{K}(N-B-1)^{k-1}{K \choose k}l+K N \alpha B} \nonumber\\
& = \frac{(N-B)^{K}}{N\sum_{k=1}^{K}(N-B-1)^{k-1}{K \choose k}} \label{Nan07}\\
  & = \frac{N-B}{N} \cdot \frac{1-\frac{1}{N-B}}{1-(\frac{1}{N-B})^{K}}, \nonumber
\end{align}
where (\ref{Nan07}) follows from the fact that $\alpha$ is a constant that does not scale with $l$, i.e., from Round 2 to Round $B$ (worst case), only one block of message is requested each round, and these download cost is negligible when the number of blocks go to infinity. 

\section{Converse}
First, we focus on the case where $2B+1  \leq N$. Consider the problem where a genie tells the user the identity of the $B$ byzantine databases.  The PIR capacity of this genie-aided case is no less than the BPIR capacity of the problem considered in Section \ref{SecSys}. Thus, the PIR capacity of the genie-aided case provides an upper bound on the capacity of the multi-round BPIR problem of interest. 

In the genie-aided scenario, due to the fact that the byzantine databases may change their answers arbitrarily, it is optimal to simply ignore the answers from the $B$ byzantine databases. This turns the problem into the classic PIR problem studied in \cite{sun2017capacity} for $N-B$ databases and $K$ messages. \cite{sun2018multiround} shows that for a classic PIR problem, compared to single-round PIR schemes, multi-round PIR scheme does not offer any performance gain in terms of the PIR capacity. Hence, the PIR capacity of the classic problem is $\frac{1-\frac{1}{N-B}}{1-(\frac{1}{N-B})^{K}}$. By multiplying the penalty term $\frac{N-B}{N}$ since all databases respond in the BPIR problem, we obtain an upper bound for the multi-round BPIR problem of interest.

Now we focus on the case where $N< 2B+1$ and $B \leq N$. We argue that in this case, the capacity of the BPIR problem is zero. Define the action $\mathcal{A}$ of a byzantine database to be: change  the realization of each of the $K$ messages to a fake realization. We see  that no matter how the user design its query, it can not distinguish between the following two cases: 
\begin{enumerate}
\item All $B$ byzantine databases perform action $\mathcal{A}$ and answer the queries with the fake realization. 
\item Only $N-B$ byzantine databases perform action $\mathcal{A}$ and answer the queries with the fake realization. 
\end{enumerate}
In both cases, the user sees $B$ databases sending one realization of the messages and $N-B$ databases sending another realization of the messages. It is impossible for the user to tell which realization is the true one.

Thus, the capacity of the BPIR problem when $2B+1 >N $ is zero. Intuitively speaking, this is the case where the number of byzantine databases is no smaller than the trustworthy databases, and hence the truth can be manipulated.

\section{Conclusion}
In this paper, we examine the connection between the RPIR and  BPIR problem, and obtain the capacity of the multi-round BPIR problem.  We show that with multi-round queries, the identities of the byzantine databases may be determined by the user, and as a result, the capacity of the BPIR problem is equal to that of the RPIR problem when the number of byzantine databases is less than half of the total number of databases. Thus, in face of byzantine databases, a multi-round query structure can indeed decrease the download cost significantly, compared to single-round queries.

\bibliography{BPIR}

\end{document}